\pdfoutput=1
\documentclass[runningheads,a4paper]{llncs}

\usepackage{amssymb}
\usepackage{cite}
\usepackage{graphicx}
\usepackage{colortbl}
\usepackage{arydshln}
\usepackage{times}
\usepackage[dvipsnames]{xcolor}
\usepackage{rotating}
\usepackage{makecell}
\usepackage{tabularx}
\usepackage{booktabs}
\usepackage{wrapfig}
\usepackage{tikz}
\usetikzlibrary{angles}
\usepackage{makecell}
\usepackage{tabu}
\usepackage{multirow}
\newcommand{\bi}{\begin{itemize}}
\newcommand{\ei}{\end{itemize}}
\newcommand{\be}{\begin{enumerate}}
\newcommand{\ee}{\end{enumerate}}
\newcommand{\fig}[1]{Figure~\ref{fig:#1}}

\newcommand{\tion}[1]{\S\ref{sect:#1}}

\usepackage{url}
\newcommand{\keywords}[1]{\par\addvspace\baselineskip
\noindent\keywordname\enspace\ignorespaces#1}
\newcommand{\crule}[3][darkgray]{\textcolor{#1}{\rule{#2}{#3}}}

\newcommand{\quartex}[4]{
\begin{picture}(25,6)%1
    {
        \color{black}
        \put(#3,3)
        {\circle*{4}}
        \put(#1,3)
        {\line(1,0){#2}}
    }
\end{picture}
}

\begin{document}

% \begin{center}
%   {\Large\bfseries\boldmath
%  Why Software Effort Estimation Needs SBSE}\\~\\
%  \normalsize{(A submission to SSBSE'18; author list blinded for review.)}
% \end{center}
\pagestyle{plain}
\thispagestyle{empty}
\title{Why Software Effort Estimation Needs SBSE}
\author{
Tianpei Xia%
\and Jianfeng Chen\and George Mathew\and Xipeng Shen\and Tim Menzies}
%\institute{blinded}
\institute{Department of Computer Science\\
North Carolina State University, NC, USA 27606\\
\url{{txia4,jchen37,george2,xshen5}@ncsu.edu, tim@menzies.us}
\authorrunning{T.Xia, J.Chen, G.Mathew,  X.Shen, T.Menzies}}
\maketitle

\begin{abstract}
Industrial practitioners now face a bewildering array of possible
configurations for effort estimation.  How to select the  best one for a particular dataset?

This paper introduces OIL (short for 
{\em \underline{o}pt\underline{i}mized 
     \underline{l}earning}),  a novel    configuration   tool for effort
estimation based on differential evolution. When tested on 945 software projects,
 OIL  significantly  improved effort estimations,
after exploring just a few configurations  (just a few    dozen). 
Further OIL's results are far better than
two    methods in widespread use: estimation-via-analogy and a
 recent state-of-the-art baseline published at TOSEM'15 by Whigham {\it et al.} 
 Given that the computational cost of this approach is so low, and the observed improvements are so large, we conclude that SBSE should be a standard component of software effort estimation.
\end{abstract}

\section{Introduction}\label{sect:intro}
This paper reports an extraordinarily successful experiment in applying
SBSE to a very common software engineering problem; i.e., effort estimation.
There are many effort estimation methods discussed
 in  the literature; e.g., 
 \bi
 \item J\"orgensen \& Shepperd report  over 250 papers
 proposing new methods for project size or effort estimation methods~\cite{jorgensen2007systematic}. 
 \item We list below 6,000+   methods
 for analogy-based effort  estimation.
 \ei
 With so many available methods, it is now
 a matter of some debate about which  one is best for a new data set.
 To simplify that task,  Whigham et al. recently proposed at TOSEM'15 a ``baseline model for software effort
 estimation'' called ATLM~\cite{Whigham:2015:BMS:2776776.2738037}. 
They recommend ATLM since, they claim, ``it performs well over a range of different project types and requires no parameter tuning''. 
Note that ``no
parameter tuning''  is an attractive property since  tuning   can be very slow--
particularly when using evolutionary genetic algorithms (GAs). For example,
the default recommendations  for GAs    suggest  $10^4$  to $10^6$     evaluations~\cite{Haupt00}. This can take some time to terminate.
Sarro et al.~\cite{sarro2016multi} reports that their evolutionary system for effort estimation
mutated 100 individuals for 250 generations. While they do not report their runtimes, we estimate that their methods would require 34 to 345 hours of CPU to terminate\footnote{Assuming 100*250 evals, 0.5 to 5 seconds to evaluate one mutation, 10-way cross-val.}.

In practice, commissioning an effort estimator on new data takes even more time than stated above. 
Wolpert's no-free lunch theorems warn that for machine learning~\cite{Wolpert96}, no single method works best on all data sets. Hence, when
building effort estimators for 
a new data set, some  {\em commissioning process} is required that tries a range of different algorithms. This is not a
mere theoretical concern:    researchers report that the ``best'' effort estimator for different
data sets   varies enormously~\cite{Minku:2011,MINKU20131512,Kocaguneli:2012}.

Given such  long runtimes, we have found it challenging to make  SBSE attractive to the broader community of  standard developers and business users. 
To address that challenge, it would be useful to have an example where  SBSE can commission a specific effort estimator for a specific data set, in just a few minutes   on a standard laptop.

This paper offers such an example. We present a surprising and fortunate result that a very ``CPU-lite'' SBSE
method  can commission an effort estimator that significantly out-performs
standard effort estimation methods. Here, by ``out-perform'' we mean that:
\bi
\item
Our estimates have statistically much smaller errors than standard methods;
\item
The comissioning time for that estimator is  very fast: median runtime for our ten-way cross-vals is just six minutes on   a   standard
8GB,  3GHz  desktop  machine.
\ei
Note that our approach is very different to much of the prior research
on effort estimation and evolutionary algorithms~\cite{BURGESS2001863,879821,5635145,5598118,Lefley:2003:UGP:1756582.1756742,sarro2017adaptive,8255666,shen02a,sarro2016multi,minku2013analysis}. Firstly, that work assumed
a ``CPU-heavy'' approach whereas we seek a ``CPU-lite'' method. Secondly, we do not defend one particular estimator; instead, our commissioning process selects   a different estimator for each data set  after exploring thousands of possibilities.

% The premise of this paper is that there are {\em too many effort
% estimation methods}. Many of those methods  exhibit
% a ``many roads lead to Rome'' property.
% That is,  when multiple  methods
%  are applied to the same data, many of them have equivalent results. 
%  For such ``many roads'' SE problems, it
%  is unnecessary to explore {\em all} those methods. Rather, it is enough
%  to intelligently sample just a small subset of those methods. 
The rest of this paper is structured as follows.
The next section describes effect estimation. We then  introduce OIL
(short for  
{\em \underline{o}pt\underline{i}mized 
     \underline{l}earning}),
 a CPU-lite search-based SE method based on differential evolution~\cite{storn1997differential}.
%  (OIL is short for 
% {\em \underline{o}pt\underline{i}mized 
%      \underline{l}earning}).
This is followed by an empirical study where estimates for 945 software projects are generated using
a variety of methods including   OIL. The results from that study  let us comment on three research questions:
\bi
\item
{\bf RQ1: Can effort estimation ignore SBSE? That is, is tuning avoidable since  just a few options are typically ``best''?} 
We will find that the ``best'' effort estimation method is highly variable.   That is, tools like OIL are important for
ensuring that the right estimators are being applied to the current data set.
\item
{\bf RQ2: Pragmatically speaking, is SBSE too hard to apply to effort estimation?}
As shown below, 
a few dozen evaluations of OIL are enough to  explore  
configuration options for   effort estimation. That is, it is hardly arduous to apply SBSE to effort
estimation. Even on a standard single core machine, the median
time to explore all those options is just  a few minutes.
\item
{\bf RQ3: Does SBSE    estimate better than widely-used effort estimation methods?}
As shown below, the  estimations from OIL perform much  better than  standard effort estimation methods, including   ATLM. 
\ei

\section{Background}\label{sect:background}
\subsection{Why Explore Software Effort Estimation?} \label{sect:see}

Software effort estimation is the process of predicting the most realistic amount of human effort (usually expressed in terms of hours, days or months of human work) required to plan, design and develop a software project based on the information collected in previous related software projects. With one or more wrong factors, the effort estimate results could be inaccurate which affect the allocated funds for the projects\cite{kemerer1987empirical}. Inadequate or overfull funds for a project could cause a considerable waste of resource and time. For example, NASA canceled its incomplete Check-out Launch Control System project after the initial \$200M estimate was exceeded by another \$200M~\cite{cowing02}. It is critical to generate effort estimations with good accuracy if for
no other reason that many government
organizations demand that the budgets allocated to
large publicly funded projects be double-checked by some estimation model~\cite{MenziesNeg:2017}.

Effort estimation can be divided into human-based techniques and model-based techniques~\cite{teak2012,shepperd2007software}.
Human-based techniques~\cite{jorgensen2004review} are that can be hard to audit or 
dispute ( e.g.,  when the estimate is generated by a senior colleague but disputed by others). Also,
empirically, it is known that humans rarely update their estimation knowledge
based on feedback from new projects~\cite{jorgensen2009impact}.

Model-based methods are preferred when   estimate have to be audited or debated
(since the  method is explicit and   available for inspection). Even
advocates of human-based methods~\cite{jorg2015a} acknowledge that model-based methods are useful for learning
the
uncertainty
about particular estimates; e.g., by running those models many times, each
time applying small mutations to the input data. 

%It can be generated using an algorithmic approach or via induced prediction systems. In the former, a senior expert proposes a general model, then use domain data to tune that model to specific projects. For example, a model named COCOMO created by Boehm~\cite{boehm1981software} hypothesized that the software development effort was exponential on LOC and linear on 15 effort multipliers such as analyst capability, product complexity, etc. Boehm defined a local calibration procedure to tune the COCOMO model to local data. If the available local training data does not conform to the requirements of a pre-defined algorithmic model like COCOMO, then the later one, induced prediction systems are useful. There are many induction methods, including linear regression, neural nets, and analogy, etc. However, all of these induction systems require a bias in order to decide what details can be safely ignored. When data violates these assumptions, finding a suitable patch could be difficult without an experienced expert.
 
  Note that this paper focuses on {\em estimation-via-analogy} and  there are many other
 ways to perform effort estimation. 
 We choose not to explore {\em parametric estimation}~\cite{MenziesNeg:2017}  since that approach
 demands 
 the data be expressed in identically the same terms as the parametric models (e.g. COCOMO).  
 This can be a major limitation to parametric models; for example,
  none of the data sets used in this paper
 are expressed in terms of the vocabulary used by standard parametric models.
 As to CPU-heavy methods (e.g.,  {\em ensembles}~\cite{Kocaguneli:2012} or  standard genetic algrithms for effort estimation~\cite{BURGESS2001863,879821,5635145,5598118,Lefley:2003:UGP:1756582.1756742,sarro2017adaptive,8255666,shen02a,sarro2016multi,minku2013analysis}),
 the   message of this paper is that  CPU-lite methods (e.g., just 40 evaluations within DE) can be surprisingly effective.  Hence, we do not explore
 CPU-heavy methods, at least for now.  
It would be interesting in future work to   check if (e.g.,) CPU-heavy ensembles or genetic algorithms are out-performed by the CPU-lite methods of this paper.

\subsection{Analogy-based Estimation (ABE)}\label{sect:abe0}
Analogy-based Estimation (ABE) was explored by Shepperd and Schofield  in 1997~\cite{shepperd1997estimating}. It is widely-used~\cite{7194627,Kocaguneli2015,7426628,6092574,MenziesNeg:2017}, in many forms.
We  say that  ``ABE0'' is the standard  form  seen in the literature
and ``ABEN'' are the 6,000+ variants of ABE  defined below. 
The general form of ABE (which applies to  ABE0 or ABEN) is: 
\bi
\item Form a table of rows of past projects. The columns of this table are composed of independent variables (the features that define projects) and one dependent variable (project  effort).
\item Find {\em training subsets.} Decide on what  similar projects (analogies) to use from the training set when examining a new test instance.
\item For each test instance, select   $k$ analogies out of the training set.
\bi
\item While selecting analogies, use a {\em similarity measure}. 
\item Before calculating similarity,  normalize   numerics  min..max to 0..1 (so all numerics   get equal chance to influence the dependent). 
\item Use {\em feature weighting} to reduce the influence of less informative features.
\ei
\item Use an {\em adaption strategy} to return some combination of  the dependent effort values seen in  the $k$ nearest analogies.
\ei
To measure    similarity between examples $x,y$, 
ABE uses $\sqrt{\sum_{i=1}^n w_i(x_i-y_i)^2}$ where $i$ ranges
over all the independent variables. In this equation, $w_i$ corresponds to feature weights applied to independent features. For ABE0, we use a uniform weighting, therefore $w_i=1$.
Also,
the {\em adaptation strategy} for ABE0 is to return the  effort values of the $k=1$ nearest analogies.
The rest of this section describes 6,000+ variants of ABE that we call ABEN.
  Note that we do \underline{{\em not}} claim that the following represents all possible ways to perform analogy-based
estimation. Rather, we merely say that (a)~all the following are
 common variations of ABE0, seen in recent research
publications~\cite{teak2012}; and (b)~anyone   with knowledge of  the current effort estimation 
literature would be tempted to try some of the following.

{\em Two ways to find training subsets}:
(1) Remove nothing: Usually, effort estimators use all training projects~\cite{chang1974finding}. Our ABE0 is using this variant;
(2) Outlier methods: prune training projects with (say) suspiciously large values~\cite{keung2008analogy}. Typically, this removes a small percentage of the training data.

{\em Eight ways to make feature weighting}:
Li {\it et al.}~\cite{li2009study} and Hall and Holmes~\cite{hall2003benchmarking} review eight different feature weighting schemes. Li {\it et al.} use a genetic algorithm to learn useful feature weights. Hall and Holmes review a variety of methods ranging from WRAPPER to various filters methods, including their preferred correlation-based method. Note that their methods assume symbolic, not numeric, dependent variables. Hence, to apply these methods we add a discretized   classes column, using (max-min)/10.  Technical aside: when we compute the errors measures (see below), we       use  the raw numeric dependent values.

{\em Three ways to discretize} (summarize numeric ranges into a few bins):
Some feature weighting schemes require an initial discretization of continuous columns. There are many discretization policies in the literature, including:
(1)~equal frequency,
(2)~equal width, 
% (4)PKID~\cite{yang2002comparative},
(3)~do nothing.

{\em Six ways to choose similarity measurements:}
Mendes {\it et al.}~\cite{mendes2003comparative} discuss three similarity measures, including the weighted Euclidean measure described above, an unweighted variant (where $w_i$ = 1), and a ``maximum distance'' measure that focuses on the single feature that maximizes interproject distance. Frank {\it et al.}~\cite{frank2002locally} offer a fourth similarity measure that uses a triangular distribution that sets to the weight to zero after the distance is more than ``k'' neighbors away from the test instance. A fifth and sixth similarity measure are the Minkowski distance measure used in~\cite{angelis2000simulation} and the mean value of the ranking of each project feature used in~\cite{walkerden1999empirical}.

{\em Four ways for adaption mechanisms:} 
(1)~median effort value,
(2)~mean dependent value,
(3)~summarize the adaptations via a second learner (e.g., linear regression)~\cite{li2009study,menzies2006selecting,baker2007hybrid,quinlan1992learning},
(4)~weighted mean~\cite{mendes2003comparative}.

{\em Six ways to select analogies:}
Kocaguneli {\it et al.}~\cite{teak2012}   says analogies selectors  are  fixed or dynamic. Fixed methods use $k\in\{1,2,3,4,5\}$
nearest neighbors
while  dynamic methods use the training set to find which $1 \le k \le N-1$ is best for   $N$ examples.\\
\begin{figure}[htp]
\centerline{\includegraphics[width=.8\textwidth]{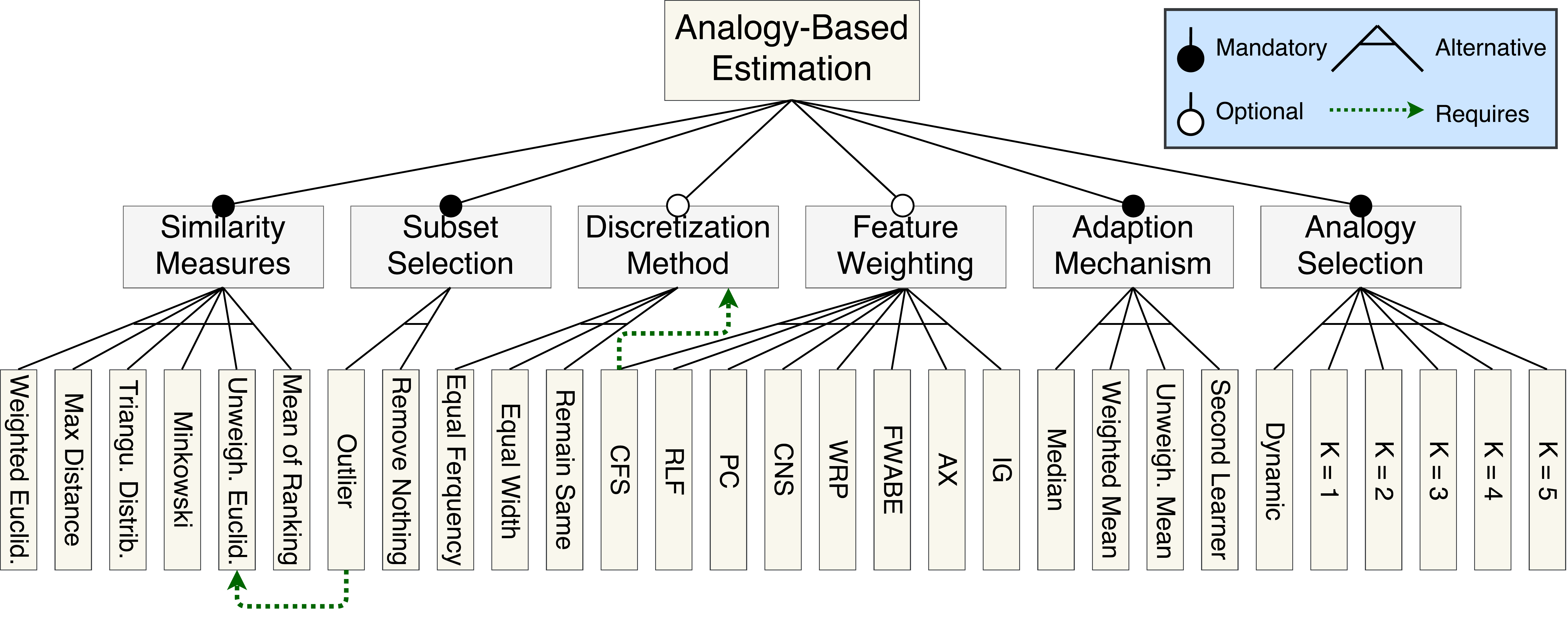}}
\caption{OIL's feature model of the space of machine learning options for ABEN.  In this model, $\mathit{Subset Selection}$, $\mathit{Similarity Measures}$, $\mathit{Adaption Mechanism}$ and $\mathit{Analogy Selection}$ are the mandatory features, while the $\mathit{Feature Weighting}$ and $\mathit{Discretization Method}$ features are optimal. To avoid making the graph too complex, some cross-tree constrains are not presented.}    
\label{fig:featuretree}
\end{figure}

\subsection{OIL}
As shown above, ABEN has
$2\times 8\times 3\times 6\times 4\times 6=6,912$ variants.  Some   can be ignored;
e.g. at $k=1$,     adaptation mechanisms return the same result, so they can be ignored. Also, not all  feature weighting techniques use discretization. But even after those discards, there are still thousands of possibilities to explore. 

OIL is  our  controller for exploring these possibilities. Initially, our plan
was to use standard hyper-parameter tuning for this task. Then we learned that (a)~standard data mining toolkits like scikit-learn lack 
some of  ABEN variants; and (b) standard hyper-parameter tuners can be  slow     (sklearn recommends a default runtime of 24 hours~\cite{sk18}).  Hence, we build OIL,  implemented as a 
 layered architecture:
 \bi
 \item
At the lowest {\em library layer}, OIL uses   Python  Scikit-Learn~\cite{pedregosa2011scikit}. 
\item
Above that, there is a {\em utilities layer} containing all the algorithms missing in Scikit-Learn (e.g., ABEN required
numerous additions at the utilities layer). 
\item
Higher up, OIL's {\em modelling layer} uses an XML-based domain-specific language to specify a feature map of data mining options.
These feature models are single-parent and-or graphs with (optionally) cross-tree constraints showing what options require or exclude other options.
A graphical representation of  the feature model used in this paper is shown in \fig{featuretree}.
\item
Finally, at top-most {\em optimizer layer}, there is some evolutionary optimizer that makes decisions across the feature map. An automatic {\em mapper} facility then links those decisions
down to the lower layers to run the selected algorithms.  
\ei
For this study, we optimize using the differential evolution method (DE~\cite{storn1997differential}), shown in \fig{DE}. 
DE was selected since certain recent software analytics papers have reported that DE can be effective for text mining~\cite{AGRAWAL2018}
and defect prediction~\cite{Fu2016TuningFS}.
While  we 
initially planned a more extensive evaluation with other optimizers,
but encountered problems accessing reference implementations\footnote{E.g. there is no reproduction package available for the Sarro et al. system~\cite{sarro2016multi} at their home page http://www0.cs.ucl.ac.uk/staff/F.Sarro/projects/CoGEE/.}.
In any case, 
the results with DE were so promising that we  deferred the application of other optimizers to future work.

  \pdfoutput=1
\begin{wrapfigure}{r}{2.55in}
\vspace{-10pt}
\scriptsize  \begin{tabular}{p{\linewidth}}
\hline
INPUT: 
\bi
\item
A dataset, as described in Table~\ref{table:dataset}; 
\item
A tuning goal $G$; e.g.,  $\mathit{MRE_{max}}$ or  $\mathit{SA_{min}}$; 
\item
DE parameters: $\mathit{np}=20$,  $\mathit{gen}=2$ or $8$, $\mathit{f}=0.8$, $\mathit{cr}=0.7$ (selected using advice from~\cite{storn1997differential}).
\ei
OUTPUT: Best tunings for learners~(e.g., ABEN) found by DE

~\\

PROCEDURE:
\bi
\item Separate the data into $\mathit{train}$ and $\mathit{tune}$;
\item Generate $np$ tunings as the initial population;
\item Score each tuning $\mathit{pop_i}$ in the population with goal $G$; 
% \item On the train data, build an ensemble and select the best:

\item  For $i=1$ to $np$  do
\be
\item Generate amutant $m$  by extrapolating  3   members of population $a$, $b$, $c$ at probability $\mathit{cr}$. For  decision $m_k \in m$:
     \bi
      \item  $m_k= a_k + f*(b_k-c_k)$ (continuous values).
      \item  $m_k= a_k \vee ~ ( b_k \vee c_k)$ (discrete values).
     \ei
\item Build a learner with parameters $m$ and train data;
\item Score $m$ on tune data using $G$;
\item Replace $\mathit{pop_i}$ with $m$ if $m$ is preferred to $\mathit{pop}_i$.;
\ee
 \item Repeat the last step until reach the number of $\mathit{gen}$;
\item Return the last population as the final result.
\ei 
\\
\hline
\end{tabular}
   \caption{OIL: uses   Storn's differential evolution method~\cite{storn1997differential}. }\label{fig:DE}
  \vspace{-20pt}
\end{wrapfigure}

  DE evolves a new generation  of candidates from a current population $\mathit{pop}_i$ of size $\mathit{np}$. Each candidate solution for effort estimation
is  pair of (Tunings, Scores) where 
Tunings are selected from the above options for ABEN;
and Scores come from training a learner using those parameters and applying it to test data.

  The premise of DE is that the best way to mutate the existing tunings is to extrapolate between current solutions. Three solutions $a, b, c$ are selected at random. For each tuning parameter $k$, at some probability $cr$, we replace the old tuning $x_k$ with $y_k$. For
 booleans $y_k = \neg x_k$ and for numerics, 
\mbox{$y_k = a_k + f \times (b_k - c_k)$}
where $f$ is a
 parameter controlling cross-over.  
The main loop of DE runs over the population, replacing old items with new candidates (if new candidate is better). This means that, as the loop progresses, the population is full of increasingly more valuable solutions (which, in turn,
helps   extrapolation).

As to the control parameters of DE,  using advice from Storn~\cite{storn1997differential}, we set $\mathit{np}=20$. 
The number of generations $\mathit{gen}\in\{2,8\}$ was set as follows. A   small number (2) was used to test
the effects of  a very   CPU-lite SBSE  effort estimator. A  larger number (8) was used to check if anything was
lost by restricting the inference to just two generations.

\section{Empirical Study} \label{sect:study}

To assess OIL, we applied it to the 945 projects
seen in nine    datasets from the SEACRAFT repository (http://tiny.cc/seacraft); see Table~\ref{table:dataset_c} and Table~\ref{table:dataset}.
This data was used since it has been  widely  used in previous estimation research.
Also, it  is quite diverse since it differs for: observation number (from 15 to 499 projects); number and type of features (from 6 to 25 features, including a variety of features describing the software projects, such as number of developers involved in the project and their experience, technologies used, size in terms of Function Points, etc.); technical characteristics (software projects developed in different programming languages and for different application domains, ranging from telecommunications to commercial information systems); geographical locations (software projects coming from China, Canada, Finland).

\pdfoutput=1
\begin{wraptable}{r}{1.7in}
\vspace{-20pt}
\scriptsize
\caption{Data   in this study. For details on the features, see Table~\ref{table:dataset}.}\label{table:dataset_c}

~\\
\centering
\begin{tabular}{r|rr}
 	&Projects&	Features\\\hline
kemerer	&15&	6\\
albrecht&	24&	7\\
isbsg10&    37& 11\\
finnish	&38&	7\\
miyazaki&	48&	7\\
maxwell&	62	&25\\
desharnais&	77&	6\\
kitchenham& 145&    6\\
china&  499&    18\\\hline
total & 945
\end{tabular}
\vspace{-18pt}
\end{wraptable}OIL collects information on two performance metrics: magnitude of the relative error (MRE)~\cite{Conte:1986:SEM:6176} and Standardized Accuracy (SA).  We make no comment
on which   measure is better-- these were selected since they are widely used in the literature.
\pdfoutput=1
\begin{table}[!b]
\caption{Descriptive Statistics of the Datasets}\label{table:dataset}
\renewcommand{\baselinestretch}{0.75} 
\centering
\begin{tabular}{cc}
\scriptsize
\begin{tabular}{|c|l|rrrr|}
    \hline
      & feature & min  & max & mean & std\\
   \hline

%%%%%%%%%%%%%%%%%%%%%%%%%%%%%%%%%%%%%%%%%%%%%%%%%%%%%%%%%%%%%%%%%%%%

\multirow{7}{*}{\begin{sideways}kemerer\end{sideways}}
& Langu. & 1 & 3 & 1.2 & 0.6\\
& Hdware & 1 & 6 & 2.3 & 1.7\\
& Duration & 5 & 31 & 14.3 & 7.5\\
& KSLOC & 39 & 450 & 186.6 & 136.8\\
& AdjFP & 100 & 2307 & 999.1 & 589.6\\
& RAWFP & 97 & 2284 & 993.9 & 597.4\\
& Effort & 23 & 1107 & 219.2 & 263.1\\
\hline
\multirow{8}{*}{\begin{sideways}albrecht\end{sideways}}
& Input & 7 & 193 & 40.2 & 36.9\\
& Output & 12 & 150 & 47.2 & 35.2\\
& Inquiry & 0 & 75 & 16.9 & 19.3\\
& File & 3 & 60 & 17.4 & 15.5\\
& FPAdj & 1 & 1 & 1.0 & 0.1\\
& RawFPs & 190 & 1902 & 638.5 & 452.7\\
& AdjFP & 199 & 1902 & 647.6 & 488.0\\
& Effort & 0 & 105 & 21.9 & 28.4\\
\hline
\multirow{12}{*}{\begin{sideways}isbsg10\end{sideways}}
% & Data\_Quality & 1 & 2 & 1.2 & 0.4\\
& UFP & 1 & 2 & 1.2 & 0.4\\
& IS & 1 & 10 & 3.2 & 3.0\\
& DP & 1 & 5 & 2.6 & 1.1\\
& LT & 1 & 3 & 1.6 & 0.8\\
& PPL & 1 & 14 & 5.1 & 4.1\\
& CA & 1 & 2 & 1.1 & 0.3\\
& FS & 44 & 1371 & 343.8 & 304.2\\
& RS & 1 & 4 & 1.7 & 0.9\\
% & Recording\_Method & 1 & 4 & 1.8 & 1.0\\
& FPS & 1 & 5 & 3.5 & 0.7\\
& Effort & 87 & 14453 & 2959 & 3518\\
\hline
\multirow{8}{*}{\begin{sideways}finnish\end{sideways}}
& hw & 1 & 3 & 1.3 & 0.6\\
& at & 1 & 5 & 2.2 & 1.5\\
& FP & 65 & 1814 & 763.6 & 510.8\\
& co & 2 & 10 & 6.3 & 2.7\\
& prod & 1 & 29 & 10.1 & 7.1\\
& lnsize & 4 & 8 & 6.4 & 0.8\\
& lneff & 6 & 10 & 8.4 & 1.2\\
& Effort & 460 & 26670 & 7678 & 7135\\
\hline

\end{tabular} 

%%%%%%%%%%%%%%%%%%%%%%%%%%%%%%%%%%%%%%%%%%%%%%%%%%%%%%%%%%%%%%%%%%%%

~
&
\scriptsize
\begin{tabular}{|c|l|rrrr|}
    \hline
      & feature
    & min  & max & mean & std\\
   \hline
%%%%%

\multirow{8}{*}{\begin{sideways}miyazaki\end{sideways}}
& KLOC & 7 & 390 & 63.4 & 71.9\\
& SCRN & 0 & 150 & 28.4 & 30.4\\
& FORM & 0 & 76 & 20.9 & 18.1\\
& FILE & 2 & 100 & 27.7 & 20.4\\
& ESCRN & 0 & 2113 & 473.0 & 514.3\\
& EFORM & 0 & 1566 & 447.1 & 389.6\\
& EFILE & 57 & 3800 & 936.6 & 709.4\\
& Effort & 6 & 340 & 55.6 & 60.1\\
\hline
\multirow{26}{*}{\begin{sideways}maxwell\end{sideways}}
& App & 1 & 5 & 2.4 & 1.0\\
& Har & 1 & 5 & 2.6 & 1.0\\
& Dba & 0 & 4 & 1.0 & 0.4\\
& Ifc & 1 & 2 & 1.9 & 0.2\\
& Source & 1 & 2 & 1.9 & 0.3\\
& Telon. & 0 & 1 & 0.2 & 0.4\\
& Nlan & 1 & 4 & 2.5 & 1.0\\
& T01 & 1 & 5 & 3.0 & 1.0\\
& T02 & 1 & 5 & 3.0 & 0.7\\
& T03 & 2 & 5 & 3.0 & 0.9\\
& T04 & 2 & 5 & 3.2 & 0.7\\
& T05 & 1 & 5 & 3.0 & 0.7\\
& T06 & 1 & 4 & 2.9 & 0.7\\
& T07 & 1 & 5 & 3.2 & 0.9\\
& T08 & 2 & 5 & 3.8 & 1.0\\
& T09 & 2 & 5 & 4.1 & 0.7\\
& T10 & 2 & 5 & 3.6 & 0.9\\
& T11 & 2 & 5 & 3.4 & 1.0\\
& T12 & 2 & 5 & 3.8 & 0.7\\
& T13 & 1 & 5 & 3.1 & 1.0\\
& T14 & 1 & 5 & 3.3 & 1.0\\
% & T15 & 1 & 5 & 3.3 & 0.7\\
& Dura. & 4 & 54 & 17.2 & 10.7\\
& Size & 48 & 3643 & 673.3 & 784.1\\
& Time & 1 & 9 & 5.6 & 2.1\\
& Effort & 583 & 63694 & 8223 & 10500\\
\hline

\end{tabular}

%%%%%%%%%%%%%%%%%%%%%%%%%%%%%%%%%%%%%%%%%%%%%%%%%%%%%%%%%%%%%%%%%%%%

~

\scriptsize
\begin{tabular}{|c|l|rrrr|}
    \hline
      & feature
    & min  & max & mean & std\\
  \hline
%%%%%

\multirow{7}{*}{\begin{sideways}desharnais\end{sideways}}
& TeamExp & 0 & 4 & 2.3 & 1.3\\
& MngExp & 0 & 7 & 2.6 & 1.5\\
& Length & 1 & 36 & 11.3 & 6.8\\
& Trans.s & 9 & 886 & 177.5 & 146.1\\
& Entities & 7 & 387 & 120.5 & 86.1\\
& AdjPts & 73 & 1127 & 298.0 & 182.3\\
& Effort & 546 & 23940 & 4834 & 4188\\
\hline
\multirow{7}{*}{\begin{sideways}kitchenham\end{sideways}}
& code & 1 & 6 & 2.1 & 0.9\\
& type & 0 & 6 & 2.4 & 0.9\\
& duration & 37 & 946 & 206.4 & 134.1\\
& fun\_pts & 15 & 18137 & 527.7 & 1522\\
& estimate & 121 & 79870 & 2856 & 6789\\
& esti\_mtd & 1 & 5 & 2.5 & 0.9\\
& Effort & 219 & 113930 & 3113 & 9598\\
\hline
\multirow{19}{*}{\begin{sideways}china\end{sideways}}
& ID & 1 & 499 & 250.0 & 144.2\\
& AFP & 9 & 17518 & 486.9 & 1059\\
& Input & 0 & 9404 & 167.1 & 486.3\\
& Output & 0 & 2455 & 113.6 & 221.3\\
& Enquiry & 0 & 952 & 61.6 & 105.4\\
& File & 0 & 2955 & 91.2 & 210.3\\
& Interface & 0 & 1572 & 24.2 & 85.0\\
& Added & 0 & 13580 & 360.4 & 829.8\\
& Changed & 0 & 5193 & 85.1 & 290.9\\
& Deleted & 0 & 2657 & 12.4 & 124.2\\
& PDR\_A & 0 & 84 & 11.8 & 12.1\\
& PDR\_U & 0 & 97 & 12.1 & 12.8\\
& NPDR\_A & 0 & 101 & 13.3 & 14.0\\
& NPDU\_U & 0 & 108 & 13.6 & 14.8\\
& Resource & 1 & 4 & 1.5 & 0.8\\
& Dev.Type & 0 & 0 & 0.0 & 0.0\\
& Duration & 1 & 84 & 8.7 & 7.3\\
& N\_effort & 31 & 54620 & 4278 & 7071\\
& Effort & 26 & 54620 & 3921 & 6481\\
\hline

\end{tabular}

%%%%%%%%%%%%%%%%%%%%%%%%%%%%%%%%%%%%%%%%%%%%%%%%%%%%%%%%%%%%%%%%%%%%

\end{tabular}
\end{table}

MRE is defined in terms of 
AR,  the magnitude of the absolute residual. This is  computed from the difference between predicted and actual effort values:
$
\mathit{AR} = |\mathit{actual}_i - \mathit{predicted}_i|
$. 
MRE is the magnitude of the relative error calculated by expressing AR as a ratio of the actual effort value; i.e., 
$
\mathit{MRE} = \frac{|\mathit{actual}_i - \mathit{predicted}_i|}{\mathit{actual}_i}
$.

MRE has been criticized~\cite{foss2003simulation,kitchenham2001accuracy,korte2008confidence,port2008comparative,shepperd2000building,stensrud2003further} as being biased towards error underestimations. 
Some researchers prefer the 
use of other (more standardized) measures, such as  Standardized Accuracy (SA)~\cite{langdon2016exact,shepperd2012evaluating}.
SA is defined in terms of 
$
\mathit{MAE}=\frac{1}{N}\sum_{i=1}^n|\mathit{RE}_i-\mathit{EE}_i|
$
where $N$ is the number of projects used for evaluating the performance, and $\mathit{RE}_i$ and $\mathit{EE}_i$ are the actual and estimated effort, respectively, for the project $i$. 
SA uses MAE as follows:
$
\mathit{SA} = (1-\frac{\mathit{MAE}_{P_{j}}}{\mathit{MAE}_{r_{guess}}})\times 100
$
where $\mathit{MAE}_{P_{j}}$ is the MAE of the approach $P_j$ being evaluated and $\mathit{MAE}_{r_{\mathit{guess}}}$ is the MAE of a large number (e.g., 1000 runs) of random guesses. 
The important thing about SA is that,
over many runs,  $\mathit{MAE}_{r_{\mathit{guess}}}$ will converge on simply using the sample mean~\cite{shepperd2012evaluating}. SA represents how much better $P_j$ is than random guessing. Values near zero means that the prediction model $P_j$ is practically useless, performing little better than  random guesses~\cite{shepperd2012evaluating}.

Note that for these evaluation measures:
\bi
\item {\em smaller} MRE values are {\em better};
\item
while {\em larger} SA values are {\em better}.
\ei
It is good practice to benchmark new methods against a variety of different approaches. Accordingly, OIL uses the following algorithms:
\bi
\item
\textbf{ABE0} was described above. It   is  widely used ~\cite{7194627,Kocaguneli2015,7426628,6092574,MenziesNeg:2017}. 
\item
\textbf{Automatically Transformed Linear Model (ATLM)}  is   an effort estimation method recently  proposed at TOSEM'15 by Whigham et al.~\cite{Whigham:2015:BMS:2776776.2738037}. ATLM is a multiple linear regression model which calculate the effort as $y_i = \beta_0 + \sum_i\beta_i\times x_{i} +  \varepsilon_i$, where $y_i$ is the $quantitative$ response for project $i$, and $x_i$ are explanatory variables. The prediction weights $\beta_i$ are determined using a least square error estimation~\cite{neter1996applied}.
Recall for the introduction that Whigham et al. recommend  ATLM since, they say, it performs well on a range of different project types and needs no parameter tuning.  
 \item
 \textbf{Differential Evolution (DE)} was  described above. Recall   we have two versions of DE.
 DE2, DE8    runs for two, eight generations  and terminate after evaluating 40, 160 configurations (respectively).
 \item
\textbf{Random Choice (RD)}. It is good practice to baseline stochastic optimizers  like DE against some
random search~\cite{nair18}. Accordingly, until it finds $N$ valid configurations, RD
selects leaves at random from \fig{featuretree}.
All these $N$ variants are executed and the best one is selected for application to the test set.   To maintain parity with   DE2 and DE8, OIL uses  $N\in\{40,160\}$ (which we denote RD40 and RD160). 
\ei
 \pdfoutput=1
\newcommand{\dbox}[1] {\crule[black!#1]{0.42cm}{0.3cm} \hspace{-0.41cm}\scalebox{1}[1.0]{{\tiny {\bf $^{#1}$}}}}

\newcommand{\wbox}[1] {\crule[black!#1]{0.42cm}{0.3cm} \hspace{-0.41cm}\scalebox{1}[1.0]{{\tiny\textcolor{white}{{\bf $^{#1}$}}}\hspace{0.04mm}}}

\begin{table}[!t]
\caption{In twenty runs of   DE2, how often was each configuration selected?  
Cells with white text denote an option selected half the time, or more. Such cells are rare. }\label{table:conf}
\begin{center}
\renewcommand{\baselinestretch}{0.75} 
\resizebox{\linewidth}{!}{
\begin{tabular}{ l|c|c|c|c|c|c}

~ & Subset & Weighting & Discret. & Similarity & Adaption  & Analogies \\\cline{2-7}

~ & 
\rotatebox[origin=c]{90}{\footnotesize
\makecell[l]{
Rm nothing\\Outlier}} &
\rotatebox[origin=c]{90}{\footnotesize
\makecell[l]{
~Remain same\\~Genetic\\~Gain rank\\~Relief\\~PCA\\~CFS\\~CNS\\~WRP
}} &
\rotatebox[origin=c]{90}{\footnotesize
\makecell[l]{
No discrete\\Equal freq.\\Equal width
}} &
\rotatebox[origin=c]{90}{\footnotesize
\makecell[l]{
~~Euclidean\\
~~Weight Euclid.\\
~~Max measure\\
~~Local likelihood~~~\\
~~Minkowski\\
~~Feature mean
}} &
\rotatebox[origin=c]{90}{\footnotesize
\makecell[l]{
~~Median\\~~Mean\\~~Second learner\\~~Weighted Mean~
}} &\rotatebox[origin=c]{90}{\footnotesize
\makecell[l]{
K=1\\
K=2\\
K=2\\
K=4\\
K=5\\
Dynamic
}} \\
\hline
%%%% auto generated from latex_configure_stats.py

kemerer&\wbox{83}\dbox{16}&\dbox{08}\dbox{06}\dbox{10}\dbox{25}\dbox{10}\dbox{06}\dbox{21}\dbox{11}&\dbox{43}\wbox{50}\dbox{06}&\dbox{28}\dbox{18}\dbox{21}\dbox{13}\dbox{11}\dbox{06}&\wbox{50}\dbox{21}\dbox{20}\dbox{08}&\dbox{01}\dbox{30}\dbox{30}\dbox{18}\dbox{08}\dbox{11}\\
albrecht&\wbox{95}\dbox{05}&\dbox{13}\dbox{05}\dbox{23}\dbox{11}\dbox{23}\dbox{11}\dbox{11}\dbox{00}&\dbox{48}\dbox{43}\dbox{08}&\dbox{28}\dbox{11}\dbox{41}\dbox{00}\dbox{18}\dbox{00}&\dbox{15}\dbox{21}\wbox{58}\dbox{05}&\dbox{10}\dbox{31}\dbox{23}\dbox{20}\dbox{11}\dbox{03}\\
isbsg10&\wbox{96}\dbox{03}&\dbox{08}\dbox{08}\dbox{45}\dbox{10}\dbox{10}\dbox{00}\dbox{15}\dbox{03}&\wbox{56}\dbox{41}\dbox{01}&\dbox{18}\dbox{25}\dbox{26}\dbox{01}\dbox{28}\dbox{00}&\dbox{30}\dbox{33}\dbox{21}\dbox{15}&\dbox{23}\dbox{30}\dbox{15}\dbox{20}\dbox{10}\dbox{01}\\
finnish&\wbox{91}\dbox{08}&\dbox{01}\dbox{04}\dbox{15}\dbox{10}\dbox{02}\dbox{13}\wbox{46}\dbox{08}&\wbox{58}\dbox{33}\dbox{08}&\dbox{26}\dbox{28}\dbox{20}\dbox{02}\dbox{23}\dbox{00}&\dbox{10}\dbox{12}\wbox{74}\dbox{03}&\dbox{06}\dbox{10}\dbox{28}\dbox{30}\dbox{23}\dbox{01}\\
miyazaki&\wbox{77}\dbox{23}&\dbox{03}\dbox{05}\dbox{12}\dbox{20}\dbox{13}\dbox{04}\dbox{27}\dbox{11}&\dbox{46}\dbox{48}\dbox{05}&\dbox{19}\dbox{20}\dbox{22}\dbox{13}\dbox{23}\dbox{01}&\dbox{26}\dbox{33}\dbox{36}\dbox{04}&\dbox{11}\dbox{21}\dbox{16}\dbox{25}\dbox{21}\dbox{04}\\
maxwell&\wbox{85}\dbox{15}&\dbox{07}\dbox{08}\dbox{24}\dbox{13}\dbox{12}\dbox{18}\dbox{14}\dbox{02}&\wbox{53}\dbox{38}\dbox{08}&\dbox{23}\dbox{26}\dbox{15}\dbox{10}\dbox{25}\dbox{00}&\dbox{18}\dbox{35}\dbox{39}\dbox{07}&\dbox{11}\dbox{16}\dbox{20}\dbox{31}\dbox{15}\dbox{06}\\
desharnais&\wbox{88}\dbox{12}&\dbox{01}\dbox{04}\dbox{20}\dbox{14}\dbox{05}\dbox{19}\dbox{28}\dbox{07}&\dbox{45}\wbox{50}\dbox{04}&\dbox{23}\dbox{21}\dbox{24}\dbox{08}\dbox{23}\dbox{00}&\dbox{39}\dbox{37}\dbox{17}\dbox{06}&\dbox{09}\dbox{15}\dbox{28}\dbox{20}\dbox{23}\dbox{03}\\
kitchenham&\wbox{63}\dbox{37}&\dbox{09}\dbox{10}\dbox{04}\dbox{13}\dbox{00}\dbox{17}\dbox{28}\dbox{17}&\dbox{47}\dbox{41}\dbox{11}&\dbox{28}\dbox{22}\dbox{19}\dbox{04}\dbox{26}\dbox{00}&\dbox{21}\dbox{28}\dbox{41}\dbox{09}&\dbox{08}\dbox{10}\dbox{20}\dbox{27}\dbox{25}\dbox{08}\\
china&\wbox{90}\dbox{09}&\dbox{02}\dbox{03}\dbox{10}\dbox{00}\dbox{07}\dbox{29}\dbox{28}\dbox{20}&\wbox{50}\dbox{45}\dbox{04}&\dbox{31}\dbox{36}\dbox{13}\dbox{00}\dbox{18}\dbox{00}&\dbox{10}\dbox{18}\wbox{67}\dbox{04}&\dbox{01}\dbox{04}\dbox{11}\dbox{27}\dbox{41}\dbox{14}\\

\end{tabular}
 }
\scriptsize{
KEY: \colorbox{black!10}{\bf 10}\colorbox{black!20}{\bf 20}\colorbox{black!30}{\bf 30}\colorbox{black!40}{\bf 40}\colorbox{black!50}{\bf \textcolor{white}{50}}\colorbox{black!60}{\bf \textcolor{white}{60}}\colorbox{black!70}{\bf \textcolor{white}{70}}\colorbox{black!80}{\bf \textcolor{white}{80}}\colorbox{black!90}{\bf \textcolor{white}{90}}\colorbox{black}{\bf \textcolor{white}{100}}\%
}
\end{center}
\end{table}
OIL performs a  $X$-fold cross validation for each of (ABE0, ATLM, DE2, DE8, RD40, RD160), for each of our nine data sets.  To apply this, datasets are partitioned into 
$X$ sets (the observations were sampled uniformly at random, without replacement),
 and then for each set OIL considered it as a testing set and the remaining 
 observations as training set. For datasets \textit{kemerer}, \textit{albrecht}, \textit{isbsg10} and \textit{finnish}, 
we uses three-fold cross validation since their instances are less than 40. For the other  larger datasets \textit{miyazaki}, \textit{maxwell}, \textit{desharnais}, \textit{kitchenham} and \textit{china},  we use ten-fold. 

Since our folds are selected in a stochastic manner, we repeat the cross-vals 20 times, each time with different random seeds.

% \begin{wrapfigure}{r}{3.5in}
% %  \begin{minipage}[c]{0.4\textwidth}
% %   \centering
% %      {\em Smaller} MRE values are {\em better}.
% %   \end{minipage}
% %   \begin{minipage}[c]{0.4\textwidth}
% %   \centering
% %      {\em Larger} SA values are {\em better}.
% %   \end{minipage}\\\noindent
% %   \begin{minipage}[c]{0.4\textwidth}
% %   \centering
% %     \includegraphics[width=\textwidth, trim={1.5cm 0 3cm 0},clip]{maxwell_mre_result.pdf}
% %     \label{fig:1}
% %   \end{minipage}
% %   \begin{minipage}[c]{0.4\textwidth}
% %     \includegraphics[width=\textwidth, trim={1.5cm 0 3cm 0},clip]{maxwell_sa_result.pdf}
% %     \label{fig:2}
% %   \end{minipage}
%   \includegraphics[width=3.5in]{maxwell.pdf}
%   \caption{Maxwell: sorted MRE and SA  values seen in 20 repeats.}\label{fig:mresa}
% \end{wrapfigure}

\section{Results}\label{sect:results}

% OIL's cross-valid experiments generate 100s of MRE and SA values for each dataset. For example, 
% Figure~\ref{fig:mresa} shows results from our six method on the   maxwell dataset. For these   plots, all  lines are sorted separately.
% As can be seen,  for maxwell,  DE8,DE2 perform very well and   ABE0 and ATLM perform much worse.

% \begin{figure}[!t]
% \scriptsize
% \begin{center}
%     \begin{minipage}{.45\line {\em Smaller} MRE values are {\em better}.
%      \includegraphics[width=2in]{Miyazaki_mre_result.pdf}
     
%      %\includegraphics[width=1.7in, trim={0.3cm 0 1cm 0},clip]{Miyazaki_mre_result.pdf}

%      {\em Larger} SA values are {\em better}.
   
%     \includegraphics[width=2in]{Miyazaki_sa_result.pdf}
%     \end{center}
    
%   \caption{MRE and SA Values of dataset miyazaki for six methods (ABE0, ALTM, DE2, DE8, RD40, RD160). For each method, the MRE values for 20 repeats are sorted.}\label{fig:mresa}
% \end{figure}

These results are divided into answers for the research questions introduced above.

~{\bf RQ1: Can effort estimation ignore SBSE? That is, is tuning avoidable since  just a few options are typically ``best''?}

 Table~\ref{table:conf} shows why SBSE is  an essential
component for effort estimation. 
This table shows how often different options were selected by the best optimizer seen in this study.
Note that, very rarely, is one option selected most of the time (exception: clearly our {\em outlier} operator is not very good-- this should be explored further in future work). 
From this table, it is clear that the   best configuration is not only data set specific, but all specific to the training set used within a data set. 
This means that {\bf RQ1=no} and tools like OIL are very important for configuring effort estimation methods.

{\bf RQ2: Pragmatically speaking, is SBSE too hard to apply to effort estimation?}

As mentioned in the introduction, some SBSE methods can be very slow. While such long runtimes 
are certainly required in other domains, for configuring effort estimation methods,  SBSE can terminate much faster than that.  Figure~\ref{fig:runtime} shows  the time required to generate our results (on a standard 8GB, 3GHz desktop machine). 
\pdfoutput=1
\begin{wrapfigure}{r}{2.25in}
\vspace{-20pt}
\caption{Mean runtime,  cross-validation (minutes), as seen in 20  repeated cross-val experiments.}\label{fig:runtime}

~\\
\setlength{\abovecaptionskip}{15pt plus 3pt minus 2pt} 
\scriptsize
\begin{tabular}{r|cccccc}

   & ~ABE0 & ~ATLM & ~RD40 & ~RD160 & ~DE2 & ~DE8 \\

\hline
kemerer   & $<$1	 & $<$1	 & 3	& 13	& 4	& 10	\\
albrecht   & $<$1	 & $<$1	 & 3	& 11	& 4	& 11	\\
isbsg10   & $<$1	 & $<$1	 & 3	& 15	& 4	& 14	\\
finnish   & $<$1	 & $<$1	 & 4	& 14	& 5	& 14	\\
miyazaki   & $<$1	 & $<$1	 & 5	& 16	& 6	& 16	\\
maxwell   & $<$1	 & $<$1	 & 12	& 52	& 18	& 53	\\
desharnais   & $<$1	 & $<$1	 & 13	& 54	& 17	& 55	\\
kitchenham   & $<$1	 & $<$1	 & 21	& 80	& 28	& 94	\\
china   & $<$1	 & $<$1	 & 57	& 232	& 52	& 243	\\

\end{tabular}
%\vspace{-15pt}
\end{wrapfigure}
  
Note that standard
effort estimation methods (i.e., ABE0 and ATLM) run very fast indeed compare to anything else. Hence, pragmatically,
it seems tempting to recommend these faster systems. Nevertheless, this paper will recommend somewhat slower methods
since, as shown below, these faster methods (i.e., ABE0 and ATLM) result in very poor estimates. 
The good news from Figure~\ref{fig:runtime} is that   cross-validation for the methods  we  will recommend (DE2) takes just a few minutes to terminate.
Hence we say that  {\bf RQ2=no} since SBSE can quite quickly commission an effort estimator, tuned specifically to a data set. 
% faster than the time required for  a cup of coffee.

% \footnote{
% According to a 2014 Reddit survey, coffee usually takes 15-20 minutes to drink, or even longer https://goo.gl/CiFzJQ.}.

%\newcommand{\nm}[1] {\hline\multicolumn{1}{c}{\cellcolor{black} { {\bf \textcolor{white}{#1}}}}}
\pdfoutput=1
\newcommand{\nm}[1] { \hline &\multicolumn{1}{l}{  {\bf #1 }}}

\newcommand{\ofr} {
{\textit{out-of-range}}
}
\newcommand{\quart}[4]{\begin{picture}(100,3)%1
{\color{black}\put(#3,3){\circle*{4}}\put(#1,3){\line(1,0){#2}}}\end{picture}}

\begin{figure}[!b]
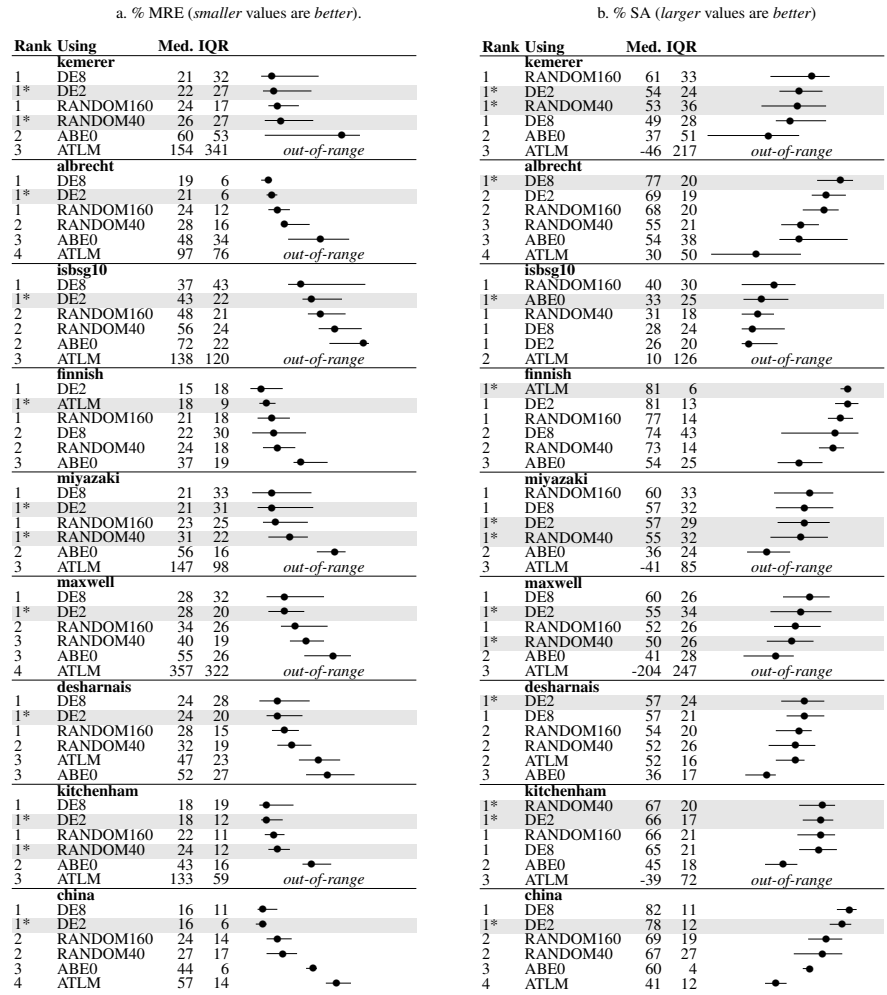

 \setlength{\belowcaptionskip}{-16pt}
 
\begin{center}
\renewcommand{\baselinestretch}{0.65} 
 \resizebox{0.49\textwidth}{!}{\begin{minipage}{3.5in}
{\small 

\begin{center}a. \% MRE  ({\em smaller} values are {\em better}).\end{center}
%\begin{tabular}{l@{~~~~}l@{~~~~}r@{~~~~}r@{~~}c@{}}
\begin{tabular}{llrrc}
  {\textbf{Rank}}& \textbf{Using} & \textbf{Med.} & \textbf{IQR} & \\

  %%% generating from latex_plotting.py::plot_mre_for_all
\nm{kemerer}\\
    1 &      DE8 &    21 &  32 & \quart{15}{32}{21}{100} \\
  \rowcolor{black!10}   1* &      DE2 &    22 &  27 & \quart{16}{27}{22}{100} \\
    1 &      RANDOM160 &    24 &  17 & \quart{19}{17}{24}{100} \\
  \rowcolor{black!10}   1* &      RANDOM40 &    26 &  27 & \quart{17}{27}{26}{100} \\
    2 &      ABE0 &    60 &  53 & \quart{17}{53}{60}{100} \\
    3 &      ATLM &    154 &  341 & \ofr \\ %\quart{96}{341}{154}{100} \\
\nm{albrecht}\\
    1 &      DE8 &    19 &  6 & \quart{15}{6}{19}{100} \\
 \rowcolor{black!10}   1* &      DE2 &    21 &  6 & \quart{18}{6}{21}{100} \\
    1 &      RANDOM160 &    24 &  12 & \quart{19}{12}{24}{100} \\
    2 &      RANDOM40 &    28 &  16 & \quart{26}{16}{28}{100} \\
    3 &      ABE0 &    48 &  34 & \quart{30}{34}{48}{100} \\
    4 &      ATLM &    97 &  76 & \ofr \\%\\\quart{59}{76}{97}{100} \\
\nm{isbsg10}\\
    1 &      DE8 &    37 &  43 & \quart{30}{43}{37}{100} \\
  \rowcolor{black!10}    1* &      DE2 &    43 &  22 & \quart{38}{22}{43}{100} \\
    2 &      RANDOM160 &    48 &  21 & \quart{41}{21}{48}{100} \\
    2 &      RANDOM40 &    56 &  24 & \quart{47}{24}{56}{100} \\
    2 &      ABE0 &    72 &  22 & \quart{53}{22}{72}{100} \\
    3 &      ATLM &    138 &  120 &  \ofr %\quart{94}{120}{138}{100}
    \\
\nm{finnish}\\
    1 &      DE2 &    15 &  18 & \quart{9}{18}{15}{100} \\
  \rowcolor{black!10}   1* &      ATLM &    18 &  9 & \quart{14}{9}{18}{100} \\
    1 &      RANDOM160 &    21 &  18 & \quart{13}{18}{21}{100} \\
    2 &      DE8 &    22 &  30 & \quart{10}{30}{22}{100} \\
    2 &      RANDOM40 &    24 &  18 & \quart{16}{18}{24}{100} \\
    3 &      ABE0 &    37 &  19 & \quart{33}{19}{37}{100} \\
\nm{miyazaki}\\
    1 &      DE8 &    21 &  33 & \quart{10}{33}{21}{100} \\
  \rowcolor{black!10}   1* &      DE2 &    21 &  31 & \quart{13}{31}{21}{100} \\
    1 &      RANDOM160 &    23 &  25 & \quart{16}{25}{23}{100} \\
  \rowcolor{black!10}   1* &      RANDOM40 &    31 &  22 & \quart{19}{22}{31}{100} \\
    2 &      ABE0 &    56 &  16 & \quart{46}{16}{56}{100} \\
    3 &      ATLM &    147 &  98 & \ofr %
    \\%\quart{91}{98}{147}{100} \\
\nm{maxwell}\\
    1 &      DE8 &    28 &  32 & \quart{18}{32}{28}{100} \\
  \rowcolor{black!10}   1* &      DE2 &    28 &  20 & \quart{19}{20}{28}{100} \\
    2 &      RANDOM160 &    34 &  26 & \quart{26}{26}{34}{100} \\
    3 &      RANDOM40 &    40 &  19 & \quart{31}{19}{40}{100} \\
    3 &      ABE0 &    55 &  26 & \quart{39}{26}{55}{100} \\
    4 &      ATLM &    357 &  322 & \ofr \\ %\quart{204}{322}{357}{100} \\
\nm{desharnais}\\
    1 &      DE8 &    24 &  28 & \quart{14}{28}{24}{100} \\
   \rowcolor{black!10}  1* &      DE2 &    24 &  20 & \quart{18}{20}{24}{100} \\
    1 &      RANDOM160 &    28 &  15 & \quart{21}{15}{28}{100} \\
    2 &      RANDOM40 &    32 &  19 & \quart{24}{19}{32}{100} \\
    3 &      ATLM &    47 &  23 & \quart{36}{23}{47}{100} \\
    3 &      ABE0 &    52 &  27 & \quart{40}{27}{52}{100} \\
\nm{kitchenham}\\
    1 &      DE8 &    18 &  19 & \quart{14}{19}{18}{100} \\
  \rowcolor{black!10}  1* &      DE2 &    18 &  12 & \quart{15}{12}{18}{100} \\
    1 &      RANDOM160 &    22 &  11 & \quart{17}{11}{22}{100} \\
  \rowcolor{black!10}   1* &      RANDOM40 &    24 &  12 & \quart{19}{12}{24}{100} \\
    2 &      ABE0 &    43 &  16 & \quart{38}{16}{43}{100} \\
    3 &      ATLM &    133 &  59 & \ofr \\%\quart{111}{59}{133}{100} \\
\nm{china}\\
    1 &      DE8 &    16 &  11 & \quart{13}{11}{16}{100} \\
  \rowcolor{black!10}   1* &      DE2 &    16 &  6 & \quart{12}{6}{16}{100} \\
    2 &      RANDOM160 &    24 &  14 & \quart{18}{14}{24}{100} \\
    2 &      RANDOM40 &    27 &  17 & \quart{18}{17}{27}{100} \\
    3 &      ABE0 &    44 &  6 & \quart{40}{6}{44}{100} \\
    4 &      ATLM &    57 &  14 & \quart{51}{14}{57}{100} \\

 %%% ----END HERE----------------------------------
  \end{tabular}} \end{minipage}}\hspace{1mm}
\resizebox{0.49\textwidth}{!}{\begin{minipage}{3.5in}
{\small   
\begin{center}b. \% SA ({\em larger} values are {\em better})\end{center}

\begin{tabular}{llrrc}
  {\textbf{Rank}}& \textbf{Using} & \textbf{Med.} & \textbf{IQR} & \\ 
   
    %%% generating from latex_plotting.py::plot_sa_for_all
\nm{kemerer}\\
    1 &      RANDOM160 &    61 &  33 & \quart{38}{33}{61}{100} \\
   \rowcolor{black!10}    1* &      DE2 &    54 &  24 & \quart{43}{24}{54}{100} \\
   \rowcolor{black!10}    1* &      RANDOM40 &    53 &  36 & \quart{33}{36}{53}{100} \\
    1 &      DE8 &    49 &  28 & \quart{41}{28}{49}{100} \\
    2 &      ABE0 &    37 &  51 & \quart{3}{51}{37}{100} \\
    3 &      ATLM &    -46 &  217 & \ofr %\quart{-231}{217}{-46}{100} 
   \\
\nm{albrecht}\\
  \rowcolor{black!10}   1* &      DE8 &    77 &  20 & \quart{64}{20}{77}{100} \\
    2 &      DE2 &    69 &  19 & \quart{61}{19}{69}{100} \\
    2 &      RANDOM160 &    68 &  20 & \quart{56}{20}{68}{100} \\
    3 &      RANDOM40 &    55 &  21 & \quart{44}{21}{55}{100} \\
    3 &      ABE0 &    54 &  38 & \quart{43}{38}{54}{100} \\
    4 &      ATLM &    30 &  50 & \quart{5}{50}{30}{100} \\
\nm{isbsg10}\\
    1 &      RANDOM160 &    40 &  30 & \quart{22}{30}{40}{100} \\
    \rowcolor{black!10}  1* &      ABE0 &    33 &  25 & \quart{23}{25}{33}{100} \\
    1 &      RANDOM40 &    31 &  18 & \quart{22}{18}{31}{100} \\
    1 &      DE8 &    28 &  24 & \quart{22}{24}{28}{100} \\
    1 &      DE2 &    26 &  20 & \quart{22}{20}{26}{100} \\
    2&      ATLM &    10 &  126 & \ofr %\quart{-104}{126}{10}{100}
  \\
\nm{finnish}\\
  \rowcolor{black!10}   1* &      ATLM &    81 &  6 & \quart{77}{6}{81}{100} \\
    1 &      DE2 &    81 &  13 & \quart{74}{13}{81}{100} \\
    1 &      RANDOM160 &    77 &  14 & \quart{70}{14}{77}{100} \\
    2 &      DE8 &    74 &  43 & \quart{44}{43}{74}{100} \\
    2 &      RANDOM40 &    73 &  14 & \quart{65}{14}{73}{100} \\
    3 &      ABE0 &    54 &  25 & \quart{42}{25}{54}{100} \\
\nm{miyazaki}\\
    1 &      RANDOM160 &    60 &  33 & \quart{40}{33}{60}{100} \\
    1 &      DE8 &    57 &  32 & \quart{41}{32}{57}{100} \\
  \rowcolor{black!10}   1* &      DE2 &    57 &  29 & \quart{42}{29}{57}{100} \\
  \rowcolor{black!10}   1* &      RANDOM40 &    55 &  32 & \quart{38}{32}{55}{100} \\
    2 &      ABE0 &    36 &  24 & \quart{25}{24}{36}{100} \\
    3 &      ATLM &    -41 &  85 & \ofr %quart{-87}{85}{-41}{100} 
    \\
\nm{maxwell}\\
    1 &      DE8 &    60 &  26 & \quart{45}{26}{60}{100} \\
    \rowcolor{black!10}   1* &      DE2 &    55 &  34 & \quart{38}{34}{55}{100} \\
    1 &      RANDOM160 &    52 &  26 & \quart{40}{26}{52}{100} \\
  \rowcolor{black!10}   1* &      RANDOM40 &    50 &  26 & \quart{36}{26}{50}{100} \\
    2 &      ABE0 &    41 &  28 & \quart{23}{28}{41}{100} \\
    3 &      ATLM &    -204 &  247 & \ofr\\%\quart{-331}{247}{-204}{100} \\
\nm{desharnais}\\
  \rowcolor{black!10}   1* &      DE2 &    57 &  24 & \quart{44}{24}{57}{100} \\
    1 &      DE8 &    57 &  21 & \quart{47}{21}{57}{100} \\
    2 &      RANDOM160 &    54 &  20 & \quart{41}{20}{54}{100} \\
    2 &      RANDOM40 &    52 &  26 & \quart{37}{26}{52}{100} \\
    2 &      ATLM &    52 &  16 & \quart{41}{16}{52}{100} \\
    3 &      ABE0 &    36 &  17 & \quart{24}{17}{36}{100} \\
\nm{kitchenham}\\
   \rowcolor{black!10}    1* &      RANDOM40 &    67 &  20 & \quart{54}{20}{67}{100} \\
   \rowcolor{black!10}    1* &      DE2 &    66 &  17 & \quart{56}{17}{66}{100} \\
    1 &      RANDOM160 &    66 &  21 & \quart{53}{21}{66}{100} \\
    1 &      DE8 &    65 &  21 & \quart{54}{21}{65}{100} \\
    2 &      ABE0 &    45 &  18 & \quart{35}{18}{45}{100} \\
    3 &      ATLM &    -39 &  72 & \ofr %\quart{-81}{72}{-39}{100}
  \\
\nm{china}\\
    1 &      DE8 &    82 &  11 & \quart{75}{11}{82}{100} \\
    \rowcolor{black!10}   1* &      DE2 &    78 &  12 & \quart{71}{12}{78}{100} \\
    2 &      RANDOM160 &    69 &  19 & \quart{59}{19}{69}{100} \\
    2 &      RANDOM40 &    67 &  27 & \quart{51}{27}{67}{100} \\
    3 &      ABE0 &    60 &  4 & \quart{56}{4}{60}{100} \\
    4 &      ATLM &    41 &  12 & \quart{35}{12}{41}{100} \\

   %%% ----END HERE----------------------------------

  \end{tabular}
  }

\end{minipage}}
\end{center}
 \caption{
\%  {\bf MRE} and \% {\bf SA} seen in 20 repeats.
 {\bf Med} is the 50th percentile and {\bf IQR} is the {\em inter-quartile range}; i.e., 75th-25th percentile. 
    Lines with a dot in the middle (e.g.,\protect\quartex{3}{13}{13}{0})
   show   median values with the IQR.   
   MRE and SA results are sorted in different directions since better MRE and SA values are smaller and larger (respectively).
   The left-hand side columns {\bf Rank} results (and the {\em smaller}, the {\em better}).
    Ranks separate statistically different results, as computed by a bootstrap test (95\% confidence)
   and the A12 test~\cite{Whigham:2015:BMS:2776776.2738037}). \ofr denote
   results that are so bad, that they  fall outside of this figure
   s range of [0,100] \%. 
   \colorbox{black!10}{1*} denotes rows of faster best-ranked methods.}
 \label{fig:jur}
\end{figure}

\newpage

{\bf RQ3: Does SBSE    estimate better than widely-used effort estimation methods?}

RQ2 showed   SBSE for effort estimation is not arduously slow. Another issue
is whether or not those SBSE methods lead to better estimates.
Figure~\ref{fig:jur} explores that issue. 
Black dots show    median values from    20 repeats.  Horizontal
lines show the  25th to 75th percentile of the values.  

The most important part of the  Figure~\ref{fig:jur} results are the {\em Rank} columns shown left-hand-side.
These ranks cluster together results that are statistically indistinguishable
as judged by a conjunction of {\em both}  a 95\% bootstrap significance test~\cite{efron93} {\em and}
a A12 test for a non-small effect size difference in the distributions~\cite{MenziesNeg:2017}. These tests were used since their non-parametric nature avoids issues with non-Gaussian
distributions.  

In Figure~\ref{fig:jur}, {\em Rank=1} denotes the better results.
When multiple treatments  receive top rank, we use the runtimes of  Figure~\ref{fig:runtime}
to break ties. 
For example, in the {\it kemerer} MRE results, four methods have {\em Rank=1}. However, 
two of these methods (DE2 and RD40) are much faster than the others.
Rows denoted \colorbox{black!10}{{\em Rank=1*}}
show these fastest top-ranked treatments.

(Technical aside: there is no statistically
significant difference between the runtimes of RD40 and DE2 in Figure~\ref{fig:runtime}, as determined by a 95\% bootstrap test.
Hence, when assigning the \colorbox{black!10}{{\em Rank=1*}}, we say that RD40 runs as fast as DE2.)

From the \colorbox{black!10}{{\em Rank=1*}} entries in  Figure~\ref{fig:jur}, we make the following comments.
\bi
\item
In marked contrast to the claims of Whigham et al., 
% ATLM is not a good effort estimation method.
ATLM does not have a very good performance.
While
it does appear as a \colorbox{black!10}{{\em Rank=1*}}  method in {\it finnish}, in all other data sets it performs badly.
Indeed, often, its performance 
% is so bad that it 
falls outside the [0,100]\% range shown in Figure~\ref{fig:jur}.\
\item
Another widely-used method in effort estimation is the ABE0 analogy-based 
effort estimator.  In 15/18 of the Figure~\ref{fig:jur}  results, ABE0 is ranked  better than ATLM.
That is, if the reader wants to avoid the added complexity of SBSE, they could ignore our advocacy for OIL and
instead just use ABE0.
That said, ABE0 is only top-ranked in 1/18 of our results. Clearly, there are better methods than ABE0.
\item
Random configuration selection performs not too badly. In 6/18 of the Figure~\ref{fig:jur}  results,
one of our random methods  earns \colorbox{black!10}{{\em Rank=1*}}. That said,  the random methods are clearly out-performed by just a few dozen evaluations of DE.
In 14/18 of these results,  DE2 (40 evaluations of DE) earns \colorbox{black!10}{{\em Rank=1*}}. 

\ei
Overall, based on the above points,  we would  recommend DE2 for comissioning effort
estimation to new data sets.
In 17/18 of our results, it gets scored   {\em Rank=1}. To be sure, in  3 of those
results, another method ran faster. However, for the sake of implementation simplicity, some researchers
might choose to ignore that minority case.
 
In summary {\bf RQ3=yes} since SBSE produces much better effort estimates than widely-used effort estimation methods.

\section{Discussion}\label{sect:discussion}

The natural question that arises from all this is why does SBSE work so well? We see  three
possibilities: (1)~DE is really clever, (2)~effort estimation is really simple, or (3)~there exists a previously undocumented {\em floor effect} in effort
estimation.

Regarding {\em DE is clever}:  DE combines  local search  (the $y=a+f*(b-c)$ extrapolation described in \fig{DE}) with an archive pruning operator (when    new candidates $y$   supplant older items in the population, then all subsequent mutations use the new and improved candidates). Hence it is wrong to characterize   40 DE evaluations as ``just 40 guesses''. Also, there is evidence from other SE domains that
DE is indeed a clever way to study SE problems. For example, Fu et al. found that hyper-parameter optimization
via a few dozen DE evaluations
was enough to produce significantly large improvements in defect prediction~\cite{Fu2016TuningFS}.
Also, in other work, Agrawal et al.~\cite{AGRAWAL2018} found that  a few
dozen evaluations of DE were enough
to significantly improve the control parameters
for the Latent Dirichlet Allocation text
mining algorithm.

Regarding {\em effort estimation  is simple}:  Perhaps  the   effective search
space of different effort estimators might be  very small. If effort estimation exhibits a ``Many roads lead to Rome'' property then
when multiple estimators are applied to the same data sets,  many of them will have equivalent performance. For such problems, configuration is not
a difficult problem since a few random probes (plus a little guidance with DE) can effectively survey all the important 
features.

Regarding {\em floor effects}: Floor effects exist when  a domain contains
some inherent performance boundary, which cannot be exceeded.
Floor effects have many causes such as the signal content of a data set is very limited,
For such data sets, then once learners reach
`the floor'', then there is no better place
to go after that.  This paper offers two pieces of evidence for floor effects in effort estimation:
\bi
\item
Recall from the above that our data sets are very small
(see Figure~\ref{table:dataset_c})-- which
suggests that effort estimation data has limited
signal. 
\item
Also, one indicator for floor effects is that informed methods perform no better
than random search and, to some extent, that indicator was seen in the above results.
Recall from the above that while a full random search was out-performed by DE2, sometimes those random
searchers performed very well indeed.
\ei
Whatever the explanation, the main effect documented by this paper   is that   a widely used SE
technique (effort estimation) which can be dramatically
improved with SBSE.

\section{Threats to Validity}\label{sect:threats}
 \textbf{Internal Bias:} All our methods contain stochastic random operators. To reduce the bias from random operators, we 
repeated our experiment in 20 times and applied statistical tests to remove spurious distinctions.

 \textbf{Parameter Bias:} DE plays an important role in OIL, in this paper, we did not discuss the influence of different DE
parameters, such as $cr$, $np$, $f$. In this paper, we followed Storn {\it et al.}'s configurations~\cite{storn1997differential}. Clearly, tuning such parameters is
a direction for  future work.

\textbf{Sampling Bias:} While we tested OIL on the nine datasets, it would be inappropriate to conclude that OIL tuning  always perform better than
others methods for all data sets.
As researchers, what we can do to mitigate this problem is to carefully document out method, release out code,
and encourage the community to try this method on more datasets, as the occasion arises.

\section{Conclusion and Future Work} \label{sect:conclusion}

This paper has explored   methods for commissioning effort estimation methods. 
As stated in the introduction, our approach is very different to much of the prior 
``CPU-heavy'' SBSE research
on effort estimation and evolutionary algorithms~\cite{BURGESS2001863,879821,5635145,5598118,Lefley:2003:UGP:1756582.1756742,sarro2017adaptive,8255666,shen02a,sarro2016multi,minku2013analysis}. Firstly, we take a  ``CPU-lite'' approach. Secondly, we do not defend one particular estimator; instead, our commissioning process selects     different estimators for different data set  after exploring thousands of options.

Our results show that SBSE is both necessary and simple to apply for effort estimation. Table~\ref{table:conf} showed that the ``best'' estimator varies greatly across effort estimation data.
Using ``CPU-lite'' SBSE methods (specifically, DE) 
it is possible to very quickly find these best estimators.
Further, the effort estimators generated by SBSE  out-perform standard methods in widespread use (ABE0 and ATLM). 
This SBSE process is not an overly burdensome task since, 
as shown above it is  enough to   perform  40 evaluations of different candidates (guided by DE).
To be sure,  some additional architecture is required for SBSE and effort estimation, but  we have packaged  that into the OIL system
(which after double blind, we will  distribute as a Python pip package).

As to future work, as  discussed in several places around this document:
\bi
\item This work should be repeated for  more     datasets. 
\item The space of operators we explored within ABEN could be expanded. Clearly, from Table~\ref{table:conf}, our {\em outliers} method is ineffective and
should be replaced. There are also other  estimation methods that could be explored (not just  for ABE, but otherwise).
\item
Other DE settings  $\mathit{np}$, $\mathit{f}$ and $\mathit{cr}$ could be explored.
\item
It could also be useful to try optimizers other than DE.
Specifically, future work could check  if (e.g.,) CPU-heavy methods such as ensembles methods~\cite{Kocaguneli:2012} or Sarro's genetic algorithms~\cite{sarro2016multi} 
are out-performed by the CPU-lite methods of this paper. That said, it should be noted that this study found no benefit in increasing the number of evaluations from  40 to 160.  Hence, possibly,  CPU-heavy
methods may not result in better estimators.
\item
It could be very insightful to explore the floor effects discussed in~\tion{discussion}. If these are very common, then that would
suggest the whole field of software effort estimation has been needlessly over-complicated.
\ei
% \section*{Acknowledgements}

% Funding source grant number blinded for review.

\tiny

\bibliographystyle{plain}
% \bibliography{reference}

\end{document}